\documentclass[letter,traditabstract,longauth]{aa}

\usepackage{graphicx}
\usepackage{txfonts}
\usepackage{natbib}

\bibpunct{(}{)}{;}{a}{}{,}

\def\mbox{\hbox}           

\def\deg{\ifmmode ^\circ                
         \else $^\circ$
         \fi
         \hskip -0.1truecm}
\def\degd#1.#2{                         
               \ifmmode {#1^{\hskip 0.05em\circ}\hskip-0.42em.\hskip0.08em#2}
               \else {#1$^{\hskip 0.05em\circ}\hskip-0.42em.\hskip0.08em$#2}
               \fi
              }
\def\mind#1.#2{                         
               \ifmmode {#1^{\hskip 0.05em\prime}\hskip-0.35em.\hskip0.05em#2}
               \else {#1$^{\hskip 0.05em\prime}\hskip-0.35em.\hskip0.05em$#2}
               \fi
              }
\def\secd#1.#2{                         
               \ifmmode {#1^{\prime\prime}\hskip-0.46em.\hskip0.12em#2}
               \else {#1$^{\prime\prime}\hskip-0.46em.\hskip0.12em$#2}
               \fi
              }
\def\timsecd#1.#2{                      
                  \ifmmode {#1^{\rm s}\hskip-0.39em.\hskip0.08em#2}
                  \else {$#1^{\rm s}\hskip-0.39em.\hskip0.08em#2$}
                  \fi
                 }
\def\hms#1h#2m#3s{                      
                  \relax
                  \ifmmode #1^{\rm h}\,#2^{\rm m}\,#3^{\rm s}
                  \else \hbox{$#1^{\rm h}\,#2^{\rm m}\,#3^{\rm s}$}
                  \fi
                 }
\def\dms#1d#2m#3s{                      
                  \relax
                  \ifmmode #1^\circ\,#2^{\prime}\,#3^{\prime\prime}
                  \else \hbox{$#1^\circ\,#2^{\prime}\,#3^{\prime\prime}$}
                  \fi
                 }
\def\dmsd#1d#2m#3.#4s{                  
                      \relax
                      \ifmmode #1^\circ\,#2^{\prime}\,#3^{\prime\prime}
                               \hskip-0.46em.\hskip0.12em#4
                      \else \hbox{$#1^\circ\,#2^{\prime}\,#3^{\prime\prime}
                            \hskip-0.46em.\hskip0.12em#4$}
                      \fi
                     }
\def\hm#1h#2m{                          
              \relax
              \ifmmode #1^{\rm h}\,#2^{\rm m}
              \else \hbox{$#1^{\rm h}\,#2^{\rm m}$}
              \fi
             }
\def\dm#1d#2m{                          
              \relax
              \ifmmode #1^\circ\,#2^{\prime}
              \else \hbox{$#1^\circ\,#2^{\prime}$}
              \fi
             }
\def\hmsd#1h#2m#3.#4s{                  
                      \relax
                      \ifmmode #1^{\rm h}\,#2^{\rm m}\,#3^{\rm s}
                               \hskip-0.39em.\hskip0.08em#4
                      \else \hbox{$#1^{\rm h}\,#2^{\rm m}\,#3^{\rm s}
                            \hskip-0.39em.\hskip0.08em#4$}
                      \fi
                     }
\def\hmd#1h#2.#3m{                  
                  \relax
                  \ifmmode #1^{\rm h}\,#2^{\rm m}
                           \hskip-0.55em.\hskip0.22em#3
                  \else \hbox{$#1^{\rm h}\,#2^{\rm m}
                        \hskip-0.55em.\hskip0.22em#3$}
                  \fi
                 }
\def\mg{\relax                          
        \ifmmode ^{\rm m}
        \else $^{\rm m}$
        \fi
       }
\def\mgd#1.#2{                          
              \relax
              \ifmmode #1^{\rm m}
                       \hskip-0.55em.\hskip0.22em#2
              \else \hbox{#1$^{\rm m}
                    \hskip-0.55em.\hskip0.22em$#2}
              \fi
             }

\def\la{\mathrel{\hbox{\rlap{\hbox{\lower4pt\hbox{$\sim$}}}\hbox{$<$}}}}
\def\ga{\mathrel{\hbox{\rlap{\hbox{\lower4pt\hbox{$\sim$}}}\hbox{$>$}}}}

\def\unitspace{\;}                      

\def\un#1{\ifmmode \unitspace\mbox{\rm #1} 
          \else $\unitspace$#1
          \fi}
\def\pun#1#2{\ifmmode \unitspace\mbox{\rm #1}^{#2} 
             \else $\unitspace$#1$^{#2}$
             \fi}
\def\per#1{\ifmmode \unitspace\mbox{\rm #1}^{-1} 
           \else $\unitspace$#1$^{-1}$
           \fi}

\def\es{\un{erg}\ps}                  
\def\esc{\es\pcmsqu}                  
\def\GHz{\un{GHz}}                    
\def\K{\un{K}}                        
\def\kms{\un{km}\pun{s}{-1}}          
\def\kpc{\un{kpc}}                    
\def\Lsun{\ifmmode \un{L}_{\odot}     
          \else $\un{L}_{\odot}$
          \fi}
\def\Mpc{\un{Mpc}}                    
\def\Msun{\ifmmode \un{M}_{\odot}     
          \else $\un{M}_{\odot}$
          \fi}
\def\muJy{\ifmmode \unitspace\mu\mbox{\rm Jy} 
          \else $\unitspace\mu$Jy
          \fi}
\def\mum{\ifmmode \unitspace\mu\mbox{\rm m} 
         \else $\unitspace\mu$m
         \fi}
\def\pc{\un{pc}}                      
\def\pcm{\per{cm}}                    
\def\pcmcub{\pun{cm}{-3}}             
\def\pcmsqu{\pun{cm}{-2}}             
\def\pyr{\pun{yr}{-1}}                
\def\ps{\pun{s}{-1}}                  
\def\sqarcsec{\ifmmode \unitspace\Box''    
              \else $\unitspace\Box''$     
              \fi} 

\def\Bp{\relax                            
        \ifmmode B_{||}                   
        \else $B_{||}$
        \fi}
\def\Bt{\relax                            
        \ifmmode B\!_{\perp}              
        \else $B\!_{\perp}$               
        \fi}
\def\Gcr{\relax                           
         \ifmmode \Gamma\!_{\rm cr}       
         \else $\Gamma\!_{\rm cr}$
         \fi}
\def\ICII{\relax                          
          \ifmmode I_{[\CII]}             
          \else $I_{[\CII]}$
          \fi}
\def\LHtwo{\relax                                 
           \ifmmode L_{\mbox{\rm\scriptsize H}_2} 
           \else $L_{\mbox{\rm\scriptsize H}_2}$  
           \fi}
\def\LIR{\relax                           
         \ifmmode L_{\rm IR}              
         \else $L_{\rm IR}$
         \fi}
\def\LLya{\relax                          
          \ifmmode L_{{\rm Ly}\,\alpha}   
          \else $L_{{\rm Ly}\,\alpha}$
          \fi}
\def\mAB{\relax                           
         \ifmmode m_{\rm AB}              
         \else $m_{\rm AB}$
         \fi}
\def\MHtwo{\relax                                 
           \ifmmode M_{\mbox{\rm\scriptsize H}_2} 
           \else $M_{\mbox{\rm\scriptsize H}_2}$  
           \fi}
\def\MHtwodot{\relax                                       
              \ifmmode \dot{M}_{\mbox{\rm\scriptsize H}_2} 
              \else $\dot{M}_{\mbox{\rm\scriptsize H}_2}$  
              \fi}                                         
\def\Mstardot{\relax                      
              \ifmmode \dot{M}_{\ast}     
              \else $\dot{M}_{\ast}$      
              \fi}
\def\nHI{\relax                                      
         \ifmmode n_{\mbox{\scriptsize\rm H\,\sc I}} 
         \else $n_{\mbox{\scriptsize\rm H\,\sc I}}$
         \fi}
\def\nHt{\relax                                
         \ifmmode n_{{\mbox{\scriptsize H}}_2} 
         \else $n_{{\mbox{\scriptsize H}}_2}$  
         \fi}
\def\rhostardot{\relax                         
                \ifmmode \dot{\rho}_{\ast}     
                \else $\dot{\rho}_{\ast}$      
                \fi}
\def\rhoZdot{\relax                          
             \ifmmode \dot{\rho}_{\rm Z}     
             \else $\dot{\rho}_{\rm Z}$      
             \fi}
\def\vhel{\relax                  %
          \ifmmode \qu{v}{hel}    
          \else $\qu{v}{hel}$     
          \fi}
\def\vLSR{\relax                  %
          \ifmmode \qu{v}{LSR}    
          \else $\qu{v}{LSR}$     
          \fi}

\def\sou#1#2{\relax                       
             \ifmmode {\rm #1}\,{\rm #2}  
             \else #1$\,$#2
             \fi}

\def\IRAS#1{\sou{IRAS}{#1}}              

\def\Mrk#1{\sou{Mrk}{#1}}                

\def\qu#1#2{\relax                          
            \ifmmode #1_{\rm #2}            
            \else $#1_{\rm #2}$
            \fi}

\def\CHp{\ifmmode \mbox{\rm CH}^+           
         \else {\rm CH}$^+$                 
         \fi}

\def\CO#1{\ifnum#1=0                    
           \ifmmode \mbox{\rm CO}
           \else {\rm CO}
           \fi
          \else
           \ifnum#1<15
            \ifmmode ^{#1}\mbox{\rm CO}
            \else $^{#1}${\rm CO}
            \fi
           \else
            \ifmmode \mbox{\rm C}^{#1}\mbox{\rm O}
            \else {\rm C}$^{#1}${\rm O}
            \fi
           \fi
          \fi}

\def\COp{\ifmmode \mbox{\rm CO}^+           
         \else {\rm CO}$^+$                 
         \fi}

\def\CS#1{\ifnum#1=0                    
           \ifmmode \mbox{\rm CS}
           \else {\rm CS}
           \fi
          \else
           \ifnum#1<15
            \ifmmode ^{#1}\mbox{\rm CS}
            \else $^{#1}${\rm CS}
            \fi
           \else
            \ifmmode \mbox{\rm C}^{#1}\mbox{\rm S}
            \else {\rm C}$^{#1}${\rm S}
            \fi
           \fi
          \fi}

\def\HCOp{\ifmmode \mbox{\rm HCO}^+          
          \else {\rm HCO}$^+$                
          \fi}

\def\Hp{\ifmmode \mbox{\rm H}^+           
        \else {\rm H}$^+$                 
        \fi}
\def\HtOp{\ifmmode \mbox{\rm H}_2\mbox{\rm O}^+   
          \else {\rm H}$_2${\rm O}$^+$            
          \fi}
\def\HthreeOp{\ifmmode \mbox{\rm H}_3\mbox{\rm O}^+   
              \else {\rm H}$_3${\rm O}$^+$            
              \fi}
\def\Hthreep{\ifmmode \mbox{\rm H}_3^+         
             \else {\rm H}$_3^+$               
             \fi}

\def\Ht{\ifmmode \mbox{\rm H}_2              
        \else {\rm H}$_2$                    
        \fi}

\def\HtO{\ifmmode \mbox{\rm H}_2\mbox{\rm O} 
         \else {\rm H}$_2${\rm O}            
         \fi}
\def\HteO{\ifmmode \mbox{\rm H}_2^{18}\mbox{\rm O} 
          \else {\rm H}$_2^{18}${\rm O}            
          \fi}

\def\OHp{\ifmmode \mbox{\rm OH}^+   
         \else {\rm OH}$^+$         
         \fi}
\def\ion#1#2{\ifmmode \mbox{{\rm #1}}\,\mbox{{\sc #2}} 
        \else {\rm #1}$\,${\sc #2}
        \fi}
\def\CI{\ion{C}{i}}
\def\CII{\ion{C}{ii}}

\def\NII{\ion{N}{ii}}
\def\OI{\ion{O}{i}}

\def\rec#1#2{\if#2a                            
              \ifmmode \mbox{{\rm #1}}\alpha   
              \else {\rm #1}$\alpha$
              \fi
             \fi
             \if#2b
              \ifmmode \mbox{{\rm #1}}\beta
              \else {\rm #1}$\beta$
              \fi
             \fi
             \if#2g
              \ifmmode \mbox{{\rm #1}}\gamma
              \else {\rm #1}$\gamma$
              \fi
             \fi}

\newcommand{\figref}[1]{Fig.~\protect\ref{#1}}

\newcommand{\eqref}[1]{Eq.~$\left(\protect\ref{#1}\right)$}

\newcommand{\secref}[1]{Sect.~\protect\ref{#1}}

\begin{document}

\title{Black hole accretion and star formation as drivers
of gas excitation and chemistry in Mrk\,231}

\author{P.P.\ van der Werf\inst{1}\thanks{\email{pvdwerf@strw.leidenuniv.nl}}
\and
K.G.\ Isaak\inst{2,3}
\and
R.\ Meijerink\inst{1}
\and
M.\ Spaans\inst{4}
\and
A.\ Rykala\inst{2}
\and
T.\ Fulton\inst{5}
\and
A.F.\ Loenen\inst{1}
\and
F.\ Walter\inst{6}
\and
A.\ Wei\ss\inst{7}
\and
L.\ Armus\inst{8}
\and
J.\ Fischer\inst{9}
\and
F.P.\ Israel\inst{1}
\and
A.I.\ Harris\inst{10}
\and
S.\ Veilleux\inst{10}
\and
C.\ Henkel\inst{7}
\and
G.\ Savini\inst{11}
\and
S.\ Lord\inst{12}
\and
H.A.\ Smith\inst{13}
\and 
E.\ Gonz\'alez-Alfonso\inst{14}
\and
D.\ Naylor\inst{15}
\and
S.\ Aalto\inst{16}
\and
V.\ Charmandaris\inst{17,29}
\and
K.M.\ Dasyra\inst{18}
\and
A.\ Evans\inst{19,20}
\and
Y.\ Gao\inst{21}
\and
T.R.\ Greve\inst{6,22}
\and
R.\ G\"usten\inst{7}
\and
C.\ Kramer\inst{23}
\and
J.\ Mart\'{\i}n-Pintado\inst{24}
\and
J.\ Mazzarella\inst{12}
\and
P.P.\ Papadopoulos\inst{25}
\and
D.B.\ Sanders\inst{26}
\and
L.\ Spinoglio\inst{27}
\and
G.\ Stacey\inst{28}
\and
C.\ Vlahakis\inst{1}
\and
M.C.\ Wiedner\inst{29}
\and
E.M.\ Xilouris\inst{30}
}

\institute{Leiden Observatory, Leiden University,
           P.O.\ Box 9513, NL-2300 RA Leiden, The Netherlands
\and
School of Physics \& Astronomy, Cardiff University, Queens Buildings, The
           Parade, Cardiff CF24 3AA, UK 
\and
ESA Astrophysics Missions Division,
           ESTEC, P.O.\ Box 299, NL-2200 AG Noordwijk, The Netherlands
\and
Kapteyn Astronomical Institute, University of Groningen, P.O.\ Box 800, NL-9700
           AV Groningen, The Netherlands
\and
Blue Sky Spectroscopy, Lethbridge, Alberta, Canada
\and
Max-Planck-Institut f\"ur Astronomie, K\"onigstuhl 17, D-69117 Heidelberg,
           Germany
\and
Max-Planck-Institut f\"ur Radioastronomie, Auf dem H\"ugel 69, D-53121 Bonn,
           Germany
\and
Spitzer Science Center, California Institute of Technology, MS 220-6,
Pasadena, CA 91125, USA
\and
Naval Research Laboratory, Remote Sensing Division, Washington, DC 20375, USA
\and
Department of Astronomy, University of Maryland, College Park, MD 20742, USA
\and
Department of Physics \& Astronomy, University College London, Gower Street,
London WC1E 6BT, United Kingdom
\and
Infrared Processing and Analysis Center,
California Institute of Technology, Pasadena, CA 91125, USA
\and
Harvard-Smithsonian Center for Astrophysics, 60 Garden Street, Cambridge, MA
           02138, USA
\and
Universidad de Alcal\'a Henares, Departamente de F\'{\i}sica, Campus
           Universitario, E-28871 Alcal\'a de Henares, madrid, Spain
\and
Department of Physics, University of Lethbridge, 4401 University Drive,
           Lethbridge, Alberta, T1J 1B1, Canada
\and
Department of Radio and Space Science,
Onsala Observatory,
Chalmers University of Technology,
SE 439 92 Onsala,
Sweden
\and
University of Crete, Department of Physics, GR-71003, Heraklion, Greece
\and
Service d'Astrophysique, CEA Saclay, Orme des Merisiers, 91191 Gif sur Yvette
           Cedex, France
\and
Department of Astronomy, University of Virginia, 530 McCormick Road,
           Charlottesville, VA 22904, USA
\and
National Radio Astronomy Observatory, 520 Edgemont Road, Charlottesville, VA
           22903, USA
\and
Purple Mountain Observatory, Chinese Academy of Sciences,
2 West Beijing Road, Nanjing 210008, China
\and
Dark Cosmology Centre,
    Niels Bohr Institute, University of Copenhagen,
    Juliane Maries Vej 30,
    2100 Copenhagen \O,
    Denmark
\and
Instituto Radioastronomie Millimetrica (IRAM), Av.\ Divina Pastora 7, Nucleo
           Central, E-18012 Granada, Spain
\and
Centro de Astrobiolog\'{\i}a (INTA-CSIC),
Ctra de Torrej\'on a Ajalvir, km 4, 28850 Torrej\'on de Ardoz,
Madrid, Spain
\and
Argelander Institut f\"ur Astronomie, Auf dem H\"ugel 71, D-53121 Bonn, Germany
\and
University of Hawaii, Institute for Astronomy, 2680 Woodlawn Drive, Honolulu, HI
           96822, USA
\and
Istituto di Fisica dello Spazio Interplanetario, CNR, Via Fosso del Cavaliere
           100, I-00133 Roma, Italy
\and
Department of Astronomy, Cornell University, Ithaca, NY 14853, USA
\and
LERMA, Observatoire de Paris, 61 Av.\ de l'Observatoire, F-75014 Paris, France
\and
Institute of Astronomy and Astrophysics, National Observatory of Athens, P.\
           Penteli, GR-15236 Athens, Greece
}

\date{Received March 31, 2010; accepted April 27, 2010}

\abstract{ We present a full high resolution SPIRE FTS spectrum of the nearby
  ultraluminous infrared galaxy $\Mrk{231}$. In total 25 lines are detected,
  including CO $J=5{-}4$ through $J=13{-}12$, 7 rotational lines of $\HtO$, 3 of
  $\OHp$ and one line each of $\HtOp$, $\CHp$, and HF\null.  We find that the
  excitation of the CO rotational levels up to $J=8$ can be accounted for by UV
  radiation from star formation.  However, the approximately flat
    luminosity distribution of the CO lines over the rotational ladder above
    $J=8$ requires the presence of a separate source of excitation for the
    highest CO lines. We explore X-ray heating by the accreting supermassive
    black hole in $\Mrk{231}$ as a source of excitation for these lines, and
    find that it can reproduce the observed luminosities.  We also consider a
    model with dense gas in a strong UV radiation field to produce the highest
    CO lines, but find that this model strongly overpredicts the hot dust mass
    in $\Mrk{231}$.  Our favoured model consists of a star forming disk of
    radius $560\pc$, containing clumps of dense gas exposed to strong UV
    radiation, dominating the emission of CO lines up to $J=8$.  X-rays from the
    accreting supermassive black hole in $\Mrk{231}$ dominate the excitation and
    chemistry of the inner disk out to a radius of $160\pc$, consistent with the
    X-ray power of the AGN in $\Mrk{231}$.  The extraordinary luminosity of the
    $\OHp$ and $\HtOp$ lines reveals the signature of X-ray driven excitation
    and chemistry in this region.}


\authorrunning{Van der Werf et al.}
\titlerunning{Seperation of AGN and starburst in $\Mrk{231}$}

\maketitle

\section{Introduction}

Carbon monoxide (CO) is a fundamental tracer of interstellar molecular gas.
However, since only the lowest 3 rotational transitions are relatively easily
accessible with ground-based telescopes, the diagnostic use of higher rotational
levels is poorly developed. This hiatus in our knowledge is becoming acute now
that high-$J$ CO observations of high-$z$ galaxies are becoming possible.

\begin{figure*}
\centering
\includegraphics[width=\textwidth]{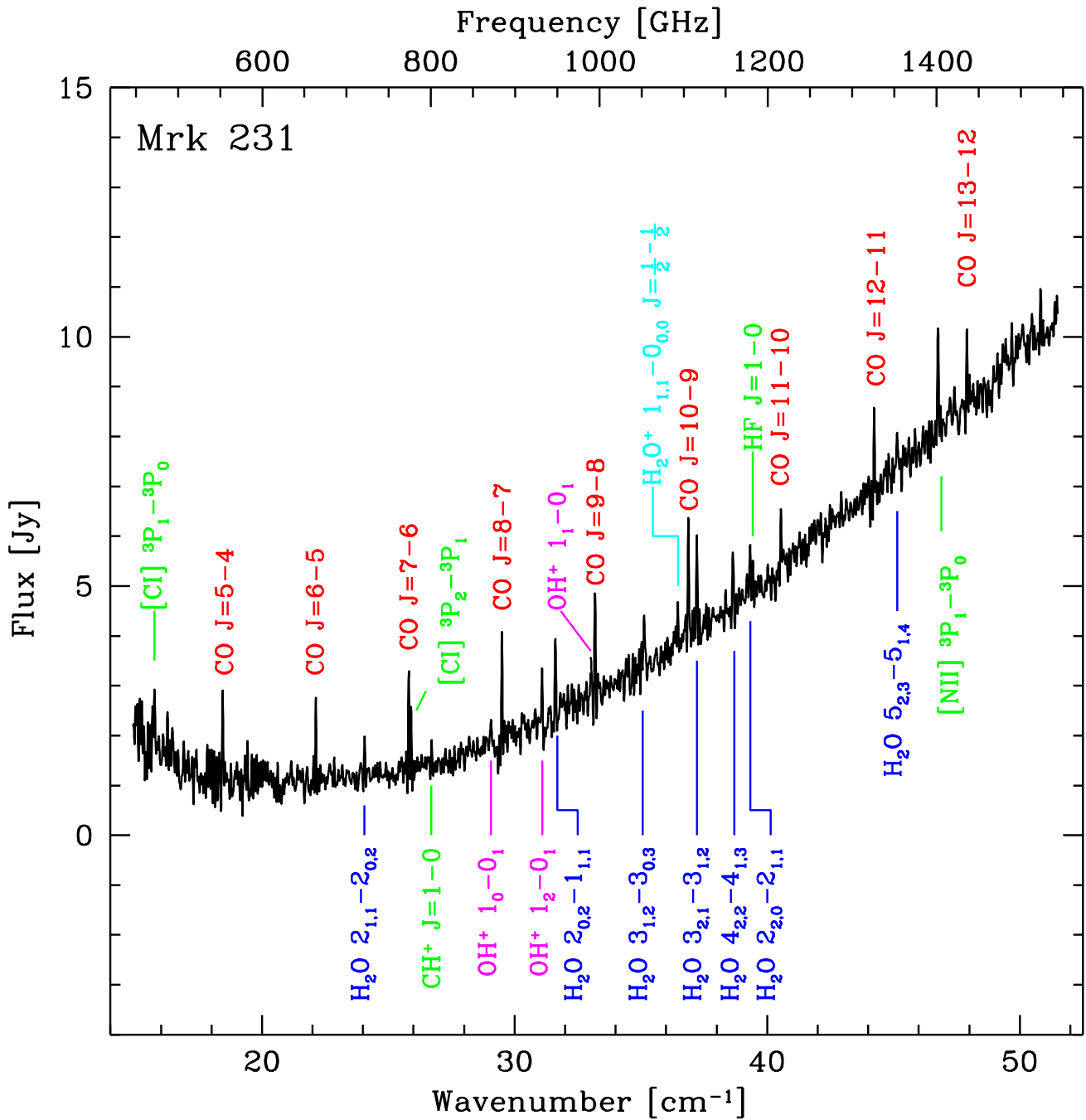}
\caption{SPIRE FTS spectrum of $\Mrk{231}$. Line identifications are given in
  red for CO lines, in blue for $\HtO$, in magenta
for $\OHp$, in cyan for $\HtOp$, and
  in green for the remaining lines.
}
\label{fig.spec}
\end{figure*}

With this in mind, we have embarked on the {\it Herschel\/} Comprehensive
(U)LIRG Emission Survey (HerCULES), an open time Key Project on the ESA {\it
  Herschel\/} Space Observatory\footnote{ {\it Herschel\/} is an ESA space
  observatory with science instruments provided by European-led Principal
  Investigator consortia and with important participation from NASA.}  (Pilbratt
et al., this issue).  The HerCULES project will establish a comprehensive
inventory of the gas cooling lines in a flux-limited sample of 29
(Ultra)luminous Infrared Galaxies or (U)LIRGs, using high spectral resolution
observations with the Fourier Transform Spectrograph (FTS) of the SPIRE
instrument (Griffin et al., this issue), combined with observations of the
[$\CII$] $158\mum$ line and the [$\OI$] 63 and $146\mum$ lines with PACS
(Poglitsch et al., this issue). Key aims of HerCULES are the development of the
diagnostic use of the gas cooling lines in local (U)LIRGs, and establishing a
local benchmark for observations of high-$z$ galaxies with the Atacama Large
Millimeter Array.  In addition, since the FTS yields full spectra, any other
luminous emission lines detected (e.g., of $\HtO$) will be available for study.

\citet{SpaansMeijerink08} have shown that X-ray excitation of the gas
  (e.g., by an AGN) and UV irradiation by young massive stars produce very
  different luminosity distributions over the CO rotational lines. Physically,
  the difference arises because X-rays penetrate a larger column density of gas
  than UV photons, and are less effective in dissociating the molecules. In
  addition, while the gas heating efficiency in a photon dominated region (PDR)
  is less than $1\%$, in X-ray dominated regions (XDRs) this efficiency is
  $10{-}50\%$.  As a result, for comparable irradiated energies, XDRs tend to
  have larger column densities of warmer molecular gas than PDRs, and will
  produce much more luminous emission in the high-$J$ CO lines. In contrast,
  PDRs are more efficient than XDRs in heating the dust. X-rays also give rise
  to significant ionization in the molecular gas and therefore drive an
  efficient ion-molecule chemistry, leading to pronounced chemical differences
  between PDRs and XDRs \citep{MeijerinkSpaans05}. Testing and using these
  diagnostics is one of the principal aims of the HerCULES project.

In this Letter, we discuss the first results of the HerCULES programme and
present the SPIRE FTS spectrum of the nearby ULIRG $\Mrk{231}$
($\sou{UGC}{8058}$, $\IRAS{F12540{+}5708}$),
the most luminous galaxy in the Revised IRAS Bright Galaxy Sample
\citep{Sandersetal03}.  Adopting $z = 0.042170$
as the heliocentric redshift of $\Mrk{231}$, correcting for the local
flow, and applying a flat 5-year WMAP
cosmology ($H_0 = 70\kms\per{Mpc}$, $\Omega_{\Lambda} = 0.73$)
yields a luminosity distance $\qu{D}{L}=192\Mpc$, with
$1''$ corresponding to $0.856\kpc$, as provided by NED\footnote{\tt
  http://nedwww.ipac.caltech.edu/}.  The derived $8-1000\mum$ luminosity of
$\Mrk{231}$ is then $\LIR=4.0\cdot10^{12}\Lsun$. $\Mrk{231}$ contains a
luminous, optically visible AGN, classified as a Seyfert~1 or a Broad Absorption
Line QSO \citep{Boksenbergetal77}.  A highly absorbed power-law
X-ray spectrum was observed by \citet{Braitoetal04} with
$\qu{L}{X}=6^{+0.6}_{-0.3}\cdot 10^{43}\un{erg}\ps$ between 2 and
$10\un{keV}$. However, $\Mrk{231}$ also
contains a kpc size disk harbouring intense star formation as shown
by high resolution radio imaging
\citep{Tayloretal99}. Interferometric imaging of CO $J=1{-}0$ and
$J=2{-}1$ emission shows an inner disk of radius $\sim520\pc$,
containing $45\%$ of the total molecular gas mass, and 
embedded in a more extended and diffuse emission component; the total molecular
gas mass is $5\cdot10^9\Msun$ 
\citep{DownesSolomon98}. The lowest part of the CO ladder (up
to $J=6{-}5$), was analysed by \citet{Papadopoulosetal07}, who
showed that the integrated CO emission can provide a
significant contribution to the total gas cooling.
Indications for X-ray-driven chemistry have been
found by \citet{Aaltoetal07} in HNC and HCN line ratios, and by
\citet{GonzalezAlfonsoetal08} in the abundances of OH and $\HtO$ observed in
absorption with ISO\null. In a comprehensive study of ULIRGs and low-$z$
quasars, \citet{Veilleuxetal09} derive a fraction of $70{\pm}15\%$ for the AGN
contribution to the far-infrared luminosity of $\Mrk{231}$, with the remainder
coming from star formation.

\section{Observations, data reduction and results}

$\Mrk{231}$ was observed in staring mode
with the SPIRE FTS on December 9, 2009,
as part of the {\it Herschel\/}
Science Demonstration Program. The high
spectral resolution mode was used, yielding
a resolution of $0.04\pcm$ over both observing bands: the
long wavelength band covering $14.9{-}33.0\pcm$ ($\lambda=671{-}303\mum$,
$\nu=467{-}989\GHz$)
and the short wavelength
band covering $32.0-51.5\pcm$ ($\lambda=313{-}194\mum$, $\nu=959{-}1544\GHz$).
In total 50 repetitions (100~FTS scans) were carried out, yielding an
on-source integration time of 6660\,s.
A reference measurement 
comprised of 120 repetitions was used to subtract the combined emission
from the sky, the telescope and instrument itself.
The data were processed and calibrated (using the asteroid
Vesta) as described in Swinyard et 
al.\ (this issue). 
Since the CO extent of $\Mrk{231}$ is at most $2''$ \citep{DownesSolomon98},
while the SPIRE beam varies from $17''$ to $42''$ over our spectrum, 
calibration procedures appropriate for a pure point source were adopted,
and no corrections for wavelength-dependent beam coupling factors were
necessary.
Because of the excellent match in the overlap region of the two spectrometer
bands ($32{-}33\pcm$), the bands were simply averaged in this region.

The full SPIRE FTS spectrum of $\Mrk{231}$ is shown in \figref{fig.spec}. It
shows a total of 25 well detected lines. The full CO ladder is detected with 9
lines from CO $J=5{-}4$ to $J=13{-}12$. In addition, 7 rotational lines of
$\HtO$ are detected, and the two [$\CI$] fine structure lines and the [$\NII$]
fine structure line, as well as rotational transitions of $\CHp$ and HF\null.
Very surprising is the detection of luminous emission from $\OHp$ and $\HtOp$.
While a possible detection of absorption in higher transitions of $\OHp$ was
reported by \citet{GonzalezAlfonsoetal08}, this is the first astronomical
detection of $\HtOp$ except in comets.

The lines are detected superposed on a continuum which rises towards the
short wavelength
side, and represents the Rayleigh-Jeans tail of the dust emission in
$\Mrk{231}$. 
A small difference
in the thermal background between the source and the reference
observation manifests itself as a moderate flux excess in the continuum at the
longest wavelengths. Line fluxes are not affected by this artefact.
For these early observations, no matching observations with the SPIRE
photometer, to measure the continuum levels, were carried out. Pending this,
quantitative analysis of the continuum is premature and in this Letter
we restrict ourselves to the spectral lines. 

Before fitting line profiles, we subtracted the continuum emission using a
grey-body fit made to the underlying spectral energy distribution (SED).
Any remaining
large-scale ripples were removed using a polynomial or sine wave fit.  Line
fluxes were recoved from this baseline-subtracted spectrum by iteratively
fitting model line profiles to this spectrum. These model line profiles are the
convolution of the FTS full resolution instrumental response (a sinc function)
with the underlying Gaussian line profile of the emission from the galaxy. 
The systematic uncertainty in the flux
scale for the lines is $20{-}30\%$ over the $21{-}52\pcm$ waveband, but
significantly higher below $21\pcm$ (which will improve when brighter
calibration sources become available). We note that the RMS fluctuations in the
spectrum are higher than the thermal noise, as a result of a fringe due to a
standing wave in the instrument, which affects the accuracy of the derived
parameters for the faintest lines in the spectrum. The removal of this
fringe, together with a search for additional faint lines, is the subject of
ongoing work.
 
\section{Discussion}

\subsection{CO excitation}
\label{sec.PDRXDR}

\begin{figure}
\resizebox{\hsize}{!}{\includegraphics{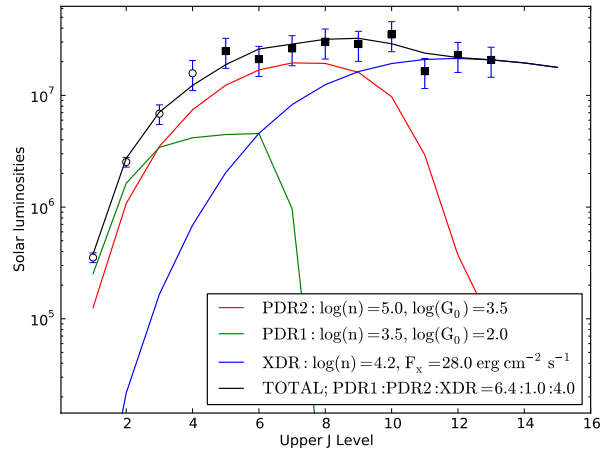}}
\caption{Luminosities of CO lines from $\Mrk{231}$. Filled symbols represent
  measurements from the SPIRE FTS spectrum, while ground-based measurements are
  denoted with open symbols. Coloured lines indicate two model PDR components
  (red and green lines) and an XDR component (blue line). The sum of these three
  components is indicated by the black line and fits the CO measurements.
  In the legend, $n$ denotes the number
  density of hydrogen nuclei ($n=\qu{n}{H}+2\nHt$) in cm$^{-3}$,
  $\qu{G}{0}$ denotes the incident UV flux in units
of $1.6\cdot10^{-3}\esc$ for the PDRs,
and $\qu{F}{X}$ the incident X-ray flux for the XDR\null. The legend also
  indicates the relative emitting areas of the three components.}
\label{fig.CO}
\end{figure}

We combine the CO line fluxes from the spectrum shown in \figref{fig.spec} with
ground-based measurements of the lower lines \citep[][and references
therein]{Papadopoulosetal07} in order to construct the CO rotational excitation
diagram shown in \figref{fig.CO}. It is seen that an approximately flat
luminosity distribution is obtained for the lines from $J=5$ upwards. Note that
the CO $J=10{-}9$ line
is blended with the $\HtO$ $3_{1,2}{-}2_{2,1}$ line which is
expected to have some luminosity ($1.8\cdot 10^7\Lsun$ for the model
by Gonz\'alez-Alfonso et al., this issue),
and this may account for its somewhat high flux. 
The total luminosity measured in the CO lines up to
$J=13{-}12$ is $(2.8\pm1.1)\cdot10^8\Lsun$.
Note that only $4\%$ of the total CO line luminosity is contained in the lowest
3 transitions. For comparison, in our Milky Way this fraction is $43\%$
  \citep{Fixsenetal99}.

The approximately flat distribution of CO line luminosity with rotational level
indicates that several excitation components must be present, since individual
components always produce a more peaked excitation diagram.
We model these components using the one dimensional
PDR/XDR models of \citet{Meijerinketal07}, as shown in \figref{fig.CO}.
The CO lines up to $J=8{-}7$ can be produced by a combination of 2 PDRs,
in qualitative agreement with the decomposition by \citet{Papadopoulosetal07}.
However, a challenge is presented by the highest CO lines:
$J=13{-}12$ and $J=12{-}11$, arising from energy levels 503 and 
$461\K$ above the ground
state. As shown in \figref{fig.CO}, these lines are strongly underproduced
by the PDRs dominating the emission in the lower lines, since the resulting gas
temperatures are not high enough for significant population of the $J>10$
levels. These lines therefore
require the presence of a third excitation component, which
can be either an XDR or a high excitation PDR\null. 

A model fit with an XDR producing the highest CO lines is shown in
\figref{fig.CO}. The required X-ray
illumination for this XDR can be produced by the AGN in
$\Mrk{231}$ \citep{Braitoetal04}, out to a
distance of $160\pc$ from the nucleus, ignoring absorption. The
ratio of radiating surfaces in the model shown in \figref{fig.CO}
implies an extended low excitation PDR
component (green curve),
with a less extended and denser central XDR region (blue
curve). Dense clouds with a smaller surface, close to massive stars and
probably embedded in the more diffuse
component, account for the medium excitation
component (red curve).

Alternatively, a very dense, high illumination PDR can account for the
highest CO lines. A good fit is found with $n=10^{6.5}\pcmcub$ and
$\qu{G}{0}=10^5$ and a surface ratio from medium to high excitation of
$1.0:0.03$.  Here the small surface area for the high excitation PDR
indicates a number of small high density clumps in a very strong UV field.
Since the radiating surface of the high excitation PDR is about $30\times$
smaller than that of the medium excitation PDR, but its density about $30\times$
larger, the $\Ht$ masses in these two components must be comparable.  For an O5
star, the required $G_0=10^5$ is reached at a radius of $0.3\pc$. For a star
formation rate of $100\Msun\pyr$ and a power law initial mass function with
slope $-2.35$ between masses of 0.3 and $120\Msun$, there are $7.6\cdot10^5$
stars of spectral type O5 or earlier in $\Mrk{231}$.  The total volume with
$G_0>10^5$ is then $8.6\cdot10^4\pun{pc}{3}$, while (for a $520\pc$ radius disk
with $15\pc$ thickness, following \citet{Daviesetal04}) the total volume of the
gas disk is {\bf $1.3\cdot10^7\pun{pc}{3}$}.  In other words, in this scenario
approximately half of the molecular mass would have to be contained in 0.7\% of
the total volume, and located within $0.3\pc$ from an O5 (or hotter) star.
  Efficient UV heating by the $G_0=10^5$ radiation field would heat the dust in
  these clumps (in total $\sim50\%$ of the total dust mass in $\Mrk{231}$) to a
  temperature of about $170\K$. In contrast, in the XDR model, where dust
  heating would be less efficient, the dust temperature would only be $\sim70\K$
  \citep{MeijerinkSpaans05}.
  
  These predictions can be tested by analysing the infrared SED of $\Mrk{231}$.
  Gonz\'alez-Alfonso et al.\ (this issue), found that the hot ($T=150{-}400\K$)
  component in the SED of $\Mrk{231}$ accounts for about $\sim20\%$ of the total
  infrared luminosity, but is produced by only 0.02\% of the total dust mass.
  This result limits the fraction of gas within $0.3\pc$ from an O5 star in
  $\Mrk{231}$ to much less than the $\sim50\%$ required to produce the highest
  CO lines with a high excitation PDR\null. This problem does not exist for the
  XDR model, which predicts most of the dust to be cooler.

\subsection{XDR chemistry}

The extraordinarily luminous emission from the molecular ions $\HtOp$ and $\OHp$
reveals the chemical signature of an XDR\null.  Assuming that the emission
  arises from a disk with $160\pc$ radius (as derived above), we can derive
  column densities in the upper levels of the relevant transitions, which
  results in values $\qu{N}{up}\sim3.0{-}3.5\cdot10^{13}\pun{cm}{-2}$ for both
  species.  Modeling the nuclear molecular gas disk in $\Mrk{231}$ with a radius
  of $520\kpc$ and $\Ht$ mass of $2.2\cdot10^9\Msun$ then results in lower
  limits to the total $\OHp$ and $\HtOp$ abundances relative to $\Ht$ of
  $\sim2\cdot10^{-10}$ in the central $160\pc$. Given the short radiative
  lifetimes ($<60\sec$) of the upper levels involved, most $\HtOp$ and $\OHp$
  molecules will be in the ground state, and total abundances will exceed these
  lower limits by large factors.  Such abundances require an efficient and
penetrative source of ionization in the molecular gas, since the production of
$\OHp$ is mainly driven by $\rm H^+ + O \rightarrow O^+ + H$ followed by $\rm
O^+ + H_2 \rightarrow OH^+ + H$ and $\rm H^+ + OH \rightarrow OH^+ + H$.
In the models considered in \secref{sec.PDRXDR}, the key species $\Hp$ is
$2{-}3$ orders of magnitude more abundant in the XDR model than in the
high excitation PDR, and the $\OHp$ abundance is larger by a comparable
amount.  A similar argument can be made for $\HtOp$, which is formed by $\rm
OH^+ + H_2 \rightarrow H_2O^+ + H$.

The extraordinary luminosity of the $\OHp$ and $\HtOp$ (and to a lesser extent
$\CHp$) lines in $\Mrk{231}$ is underlined by a comparison with the SPIRE
spectrum of the Orion bar PDR (Habart et al., this issue),
which shows no trace of
$\OHp$ or $\HtOp$, and only weak $\CHp$ emission, while in $\Mrk{231}$ the lines
are only a factor $2{-}3$ fainter than the CO lines.
While enhanced cosmic ray fluxes in a starburst environment
will increase the degree of ionization and hence the
production of $\OHp$ and $\HtOp$ in a PDR, they do not elevate the temperatures
to the level required to produce the highest CO lines \citep{Meijerinketal06}.
It is thus the combination of strong high-$J$ CO lines and high $\OHp$ and
$\HtOp$ abundances that reveals X-ray driven excitation and
chemistry in $\Mrk{231}$.

\section{Outlook}

We have shown that the SPIRE spectrum of $\Mrk{231}$ reveals both PDR and
XDR emission lines, and made a separation of these components.
A key goal of the HerCULES project will be using this decomposition for a
quantitative separation between star formation and black
hole accretion as power sources for the infrared luminosities of
dusty galaxies. In the case of $\Mrk{231}$, this issue will be addressed in
a forthcoming paper (Meijerink et al., in prep.). Data
will be obtained as part of the HerCULES program for an
additional 28 objects, which will enable us to put the results presented
here on a statistically significant footing.

\begin{acknowledgements}
We thank Ewine van Dishoeck, Xander Tielens, and Thomas
Nikola for useful discussions. We especially thank Ed Polehampton, Peter
Imhof-Davies and Bruce Swinyard for their help with the FTS data processing.
JF thanks MPE for its hospitality. The Dark Cosmology Centre is funded by the
DNRF\null.
The following institutes have provided hardware and software elements to the
  SPIRE project: University of Lethbridge, Canada; NAOC, Beijing, China; CEA
  Saclay, CEA Grenoble and LAM in France; IFSI, Rome, and University of Padua,
  Italy; IAC, Tenerife, Spain; Stockholm Observatory, Sweden; Cardiff
  University, Imperial College London, UCL-MSSL, STFC-RAL, UK ATC Edinburgh, and
  the University of Sussex in the UK\null. 
Funding for SPIRE has been provided by
  the national agencies of the participating countries and by internal institute
  funding: CSA in Canada; NAOC in China; CNES, CNRS, and CEA in France; ASI in
  Italy; MEC in Spain; Stockholm Observatory in Sweden; STFC in the UK; and NASA
  in the USA.  Additional funding support for some instrument activities has
  been provided by ESA\null.
\end{acknowledgements}

\bibliographystyle{aa}

\bibliography{%
strings,%
Arp220,%
cosmology,%
ISM,%
Mrk231,%
photodissociation,%
ULIRGs,%
XDRs%
}

\end{document}